\documentclass{article}
\usepackage{spconf,amsmath,graphicx,tabularray,makecell,cite, xfp}
\usepackage[table]{xcolor}
\usepackage{amssymb}
\usepackage{hyperref}

\newcommand{\showgap}[2]{%
\ifdim #2pt > #1pt 
  \textcolor{green!75!black}{%
    \fpeval{round(#2 - #1, 2)}%
    \,$\uparrow$}%
\else 
  \textcolor{red!85!black}{%
    \fpeval{round(#1 - #2, 2)}%
    \,$\downarrow$}%
\fi
}

\title{LibEMER: A novel benchmark and algorithms library for EEG-based Multimodal Emotion Recognition}
\name{Zejun Liu$^{\star}$ \qquad Yunshan Chen$^{\dagger}$ \qquad Chengxi Xie$^{\star}$ \qquad Yugui Xie$^{\bigstar}$ \qquad Huan Liu$^{\dagger\#}$\thanks{$^{\#}$ corresponding author.}}
\address{$^{\star}$ XJTU-POLIMI Joint School, Xi'an Jiaotong University \\ $^{\dagger}$ School of Computer Science and Technology, Xi'an Jiaotong University \\ $^{\bigstar}$ MIGU Video Co., Ltd., Shanghai, China} 
\begin{document}
\ninept
\maketitle
\begin{abstract}
EEG-based multimodal emotion recognition(EMER) has gained significant attention and witnessed notable advancements,
the inherent complexity of human neural systems has motivated substantial efforts toward multimodal approaches.
However, this field currently suffers from three critical limitations: (i) the absence of open-source implementations.
(ii) the lack of standardized and transparent benchmarks for fair performance analysis.
(iii) in-depth discussion regarding main challenges and promising research directions is a notable scarcity.
To address these challenges, we introduce LibEMER, a unified evaluation framework 
that provides fully reproducible PyTorch implementations of curated deep learning methods alongside standardized protocols 
for data preprocessing, model realization, and experimental setups. 
This framework enables unbiased performance assessment on three widely-used public datasets 
across two learning tasks. 
The open-source library is publicly accessible at: \href{https://anonymous.4open.science/r/2025ULUIUBUEUMUEUR485384}{\underline{LibEMER}}
\end{abstract}
\section{Introduction}
\label{sec:intro}

EEG-based multimodal emotion recognition (EMER) represents a critical research domain within affective computing, 
focusing on the development of computational models for precise identification of 
human emotional states. As electroencephalography (EEG) provides direct measurements 
of cortical neural activity, it has received continuous attention. 
Furthermore, given the inherent complexity of human neurophysiological systems,
multimodal approaches integrating EEG with complementary physiological signals 
have increasingly emerged as a key research focus in recent years
These supplementary modalities encompass: electrocardiography(ECG), 
blood volume pressure(BVP), respiration(RSP), electrooculography(EOG), 
electromyography(EMG), galvanic skin response(GSR), eye movements and facial expressions. 

In the field of EER, TorchEEGEMO\cite{zhang2024torcheegemo} provides a deep learning toolkit, GNN4EEG\cite{zhang2024gnn4eeg} focuses on evaluating SOTA GNN-based methods.
However, neither establishes a sufficiently rigorous framework for benchmarking, thus failing 
to address the prevailing lack of standardized benchmark. While LibEER\cite{liu2025libeer} presents a more comprehensive and rigorous benchmark 
with deeper analysis, its applicability is confined to unimodal EEG-based approaches, rendering it unsuitable 
for multimodal scenarios that integrate EEG with other physiological signals. In contrast, 
our work establishes a rigorous benchmark for multimodal scenarios.

Although substantial progress has been made in developing deep learning models for EMER, some systematic issues that hinder development of relevant research remain unsolved.
The most core challenge is the absence of comprehensive standardized benchmarks, which fosters significant discrepancies 
in preprocessing pipelines and experimental configurations, thereby compromising transparency and comparability across studies. The problem is compounded 
by insufficience of details in publications and open-source codebases, which collectively obstruct verification and 
replication efforts. Even when methods are open-sourced, they are often implemented on different platforms like PyTorch\cite{paszke2019pytorch} and TensorFlow\cite{abadi2016tensorflow}, 
and their training processes also have various detailed discrepancies, making a fair and unified evaluation difficult.Furthermore, while existing surveys\cite{liu2024eeg,pillalamarri2025review} have summarized the achivements, 
they generally fall short of providing a critical evaluation and insights to inspire promising progress. 
To address these challenges, we introduce LibEMER, a benchmark and open-source 
algorithm library based on PyTorch that establishes unified standards 
for model implementation, data processing, experimental configurations, 
and evaluation metrics. This toolkit provides fully reproducible workflows and supports 
three widely-adopted public datasets, SEED\cite{zheng2015investigating}, SEEDV\cite{li2019classification}, and DEAP\cite{koelstra2011deap}, 
across two critical learning paradigms: subject-dependent and 
subject-independent recognition tasks. Comprehensive evaluation 
analyses are conducted on ten state-of-the-art methods 
across multiple datasets under standardized benchmark.

Overall, our work makes the following key contributions:
\begin{itemize}
    \item We introduce an open-source multimodal deep learning library for emotion recognition 
    based on the fusion of EEG and other physiological signals. Our library implements a standardized, 
    end-to-end workflow, containing every stage from data loading to training and evaluation. It facilitates 
    efficient adoption, flexible configuration, and streamlined experimentation for researchers.

    \item We establish a fair, transparent, and standardlized benchmark for  multimodal EEG-based emotion recognition. 
    This benchmark provides meticulously defined protocols for data preprocessing, experimental tasks, and evaluation metrics, 
    thereby ensuring an unbiased and consistent assessment of diverse methods.

    \item We conduct extensive experiments on the implemented methods, strictly adhering to the proposed 
    benchmark, and provide a thorough analysis of the results. We systematically report the performance of various models across different tasks and datasets, 
    offering valuable insights and actionable recommendations that can guide and inspire future research 
    in the field.
\end{itemize}

\section{Benchmark Building}
\label{sec:benchmark}
In EEG-based multimodal emotion recognition, no open, transparent and 
well-organized benchmark exists. In this section, we delineate
methods selection, datasets, data loading and preprocessing, data splitting,
and evaluation metrics. Table\ref{tab:benchmark} shows the key information of proposed benchmark. 

\begin{table}
  \centering
  \footnotesize
  \setlength\tabcolsep{2.5pt} 
  \caption{key modules and information of proposed benchmark}
  \label{tab:benchmark}
  \begin{tblr}{
    width = \columnwidth,
    column{2} = {c},
    hlines,
    hline{2} = {-}{0.08em},
    rowsep = 1pt
  }
  Module                  & Key Information                                                             \\
  Dataset                 & SEED/SEEDV/DEAP                                                             \\
  EEG           & {Baseline Removal/ Bandpass filter/Artifacts \\Removal/DE Features/Segment} \\
  Eye movement  & Extracted Features /Segment                                                \\
  PPS           & Segment/ Statical features                                                  \\
  Data Split              & Train : Val : Test = 3 : 1 : 1                                              \\
  Tasks                   & Subject-dependent/Subject-independent                                       \\
  Metrics                 & Accuracy/F1-Score                                                           
  \end{tblr}
  \end{table}

\subsection{Methods Selection}
\label{ssec:methods selections}

This study adopts a systematic methodology curation approach 
based on rigorous review of recent publications in leading 
domain-specific venues (such as TAFFC, ACM MM, ICONIP). 
Baseline methods were extracted and frequency-ranked to 
identify highly-cited recurrent techniques, supplemented by 
SOTA methods published since 2023. From the resulting ranked list, 
21 representative methods were selected and categorized 
by architectural paradigm: DNN, CNN, RNN, Transformer, 
and GNN. Each method underwent rigorous reimplementation with
10 \% replication error tolerance. The final curated 
collection of ten methods ensures minimum representation
per architectural category.

{\bf DNN:} DNNs are forward fully-connected networks that learn deep emotional representations 
by mapping physiological signals into a high-dimensional space. The library contains DCCA\cite{qiu2018multi}, DCCA\_AM\cite{liu2021comparing}
,BDDAE\cite{tang2017multimodal}, CFDA-CSF\cite{jimenez2024cfda}. 

{\bf CNN:} CNNs use convolutional operations to extract local patterns. In EMER, 
they typically process multi-channel signals as 2D representations, such as topological electrode maps. 
G2G\cite{jin2023graph} and CRNN\cite{liao2020multimodal} are included.

{\bf RNN:} RNNs utilize recurrent connections to model temporal dependencies in 
sequential data. In EMER, they are primarily applied to capture signal dynamics over time, 
though some also process channels sequentially to extract spatial features.
BimodalLSTM\cite{tang2017multimodal} is a representative algorithm. CRNN also cotains RNN architecture.

{\bf Transformer:} Transformer employs a self-attention mechanism to capture long-range dependencies.
For EMER, signals are segmented into patches (by time, channel, or region) to learn their contextual interrelations.
CMCM\cite{zhang2024cross} and MCAF\cite{yin2024research} are incorporated.

{\bf GNN:} GNNs are designed for graph-structured data, making them ideal for modeling the 
non-Euclidean nature in brain connectivity. In this approach, EEG channels are treated as 
nodes, and their relationships are captured by an adjacency matrix that can be either predefined 
from spatial proximity or learned adaptively. HetEmotionNet\cite{jia2021hetemotionnet} is incorporated.

\subsection{Datasets}
\label{ssec:datasets}

Through systematic evaluation of usage frequency, 
sample scale, data collection and modality coverage we select SEED, SEEDV, and DEAP 
as our primary evaluation corpora. Table\ref{tab:dataset} shows the basic information.

{\bf SEED\cite{zheng2015investigating}:} SEED contains EEG recordings from 15 adults (8 male, 7 female), 
though eye movement data exists only for 12 subjects. 
Each participant completed three experimental sessions, viewing 
15 video stimuli per session corresponding to three emotional 
states (positive, neutral, negative). It delivers 
raw data, PSD features, and DE features of EEG data. For eye movement, it provides 
excel files of eye tracking information and extracted features.

{\bf SEEDV\cite{li2019classification}:} The SEEDV dataset Comprises EEG and eye movement data from 16 participants
(6 male, 10 female), this dataset captures three experimental sessions per subject.
Each session presents 15 validated video stimuli eliciting 5 emotional states 
(happy, sad, neutral, disgust, fear). EEG data contains both raw data and DE featrues.
concurrently, eye movement data encompasses information in tabular format and extracted features.

{\bf DEAP\cite{koelstra2011deap}:} The DEAP dataset encompasses multimodal physiological signals from 32 participants 
viewing 40 standardized one-minute music video stimuli, with 
self-reported assessments across four dimensions (arousal, valence, dominance, and liking) 
following each stimulus presentation; It provides both the original signals 
sampled at 512 Hz and a preprocessed 40-channel downsampled version (128 Hz) 
integrating 32 EEG channels alongside 8 PPS channels.

\begin{table}
  \centering
  \caption{The basic information of datasets in LibEMER. }
  \scriptsize
  
  \label{tab:dataset}
  \begin{tblr}{
    column{even} = {c},
    column{3} = {c},
    column{5} = {c},
    hline{1-2,5} = {-}{},
    colspec = {c c c c X[c,m] X[c,m]},
    rowsep= 1pt,
    colsep= 3pt,
  }
  Dataset & Session & Subject & Trial & Modality                          & Labels                             \\
  SEED    & 3       & 12      & 15    & EEG/Eye movement                  & {positive/neutral/\\negative}      \\
  SEEDV   & 3       & 16      & 15    & EEG/Eye movement                  & {happy/sad/fear/\\disgust/neutral} \\
  DEAP    & 1       & 32      & 40    & {EEG/EOG/EMG/\\GSR/RSP/\\TEMP/BVP} & arousal/valence                    
  \end{tblr}
  \end{table}

\subsection{Data Preprocessing}
\label{ssec:preprocess}

A significant challenge in multimodal emotion recognition lies in the absence of 
unified and standardized preprocessing protocols. To address this gap, we establish
a preprocessing framework tailored to EEG characteristics and dataset modality properties. 
For EEG processing: baseline removal is first performed, followed by 0.3-50 Hz band-pass filtering. 
Artifact contamination is subsequently mitigated through principal component analysis 
(PCA). We then extract differential entropy (DE) features\cite{duan2013differential} across five canonical 
frequency bands($\delta$, $\theta$, $\alpha$, $\beta$, $\gamma$)\cite{sanei2013eeg} and apply linear dynamic 
system (LDS) to smooth the features\cite{duan2012eeg}.

For the \textbf{SEED} and \textbf{SEEDV} datasets, eye movement is utilized as a supplementary modality.
From this data, a total of 33 features\cite{lu2015combining} are extracted using a 4s non-overlapping window.
These features included the Differential Entropy (DE) of pupil diameters and dispersion 
on the X and Y axes, fixation duration, blink duration, and additional statistical features 
such as blink frequency and saccade frequency. To ensure temporal alignment between 
EEG and eyemovement, the DE features from the EEG signals were similarly extracted 
using a 4s non-overlapping window.

For the \textbf{DEAP} dataset, since previous research lacks a standardized approach for 
extraction of peripheral physiological signals, with methods varying between 
the use of raw signals and various statistical features. In this work, we adopt a 
unified methodology, extracting EEG DE features and peripheral physiological signal 
data with a one-second non-overlapping window. For methodologies that originally 
specified the use of raw signals, we maintained this approach. However, for methods 
that utilized statistical features or had not been previously tested on the DEAP 
dataset, a feature set was established through a voting method\cite{qiu2018multi}\cite{liu2021comparing}\cite{tang2017multimodal}. 
This process involved computing the maximum value, minimum value, mean, standard 
deviation, variance, and sum of squares for each of the 8 channels, yielding a 
48-dimensional feature in total.
\subsection{Data Splitting}
\label{ssec:split}
The strategy for dataset spllitting critically impacts the performance of model. We observe
that most studies directly split data into train and test sets and report peak 
performance on test set without setting up validation set for model selection.
Although no data leakage occurs, the approach can raise risk of implicit model tuning
towards test set, which may lead to overestimated performance and fail to reflect the 
true generalization ability of model.  

To address this issue, we adopt a strict cross-validation strategy. Data will be splitted
into train, validation and test in a 3:1:1 ratio, with no overlap between them. Model selection is based on optimal
performance on validation set. Final reported performance is then derived from evaluation of
the selected model on test set.The splitting strategy yields a more accurate 
assessment of the model's true generalization and performance across tasks.

\begin{figure}[htb]
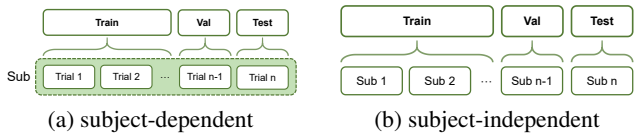

   \begin{minipage}[b]{0.48\linewidth}
   \centering
  \centerline{\includegraphics[width=4.0cm]{figures/sub\_dependent\_2.pdf}}
   \centerline{(a) subject-dependent}\medskip
   \end{minipage}
   \hfill
   \begin{minipage}[b]{0.48\linewidth}
   \centering
   \centerline{\includegraphics[width=4.0cm]{figures/sub\_independent\_2.pdf}}
   \centerline{(b) subject-independent}\medskip
   \end{minipage}
   \caption{Two primary tasks for emotion recognition}
   \label{fig:task}
   \end{figure}

\subsection{Evaluation and Metrics}
\label{ssec:metrics}

We evaluate the model using two primary emotion recognition tasks shown in Figure\ref{fig:task}: subject-dependent 
and subject-independent evaluation. In the \textbf{Subject-Dependent (SD)} task, 
the training and test sets are derived from different trials of the same subject in 
the same session. This task primarily assesses the model's performance in scenarios 
requiring personalization and customization. In the \textbf{Subject-Independent (SI)} task, 
the training and test data are partitioned based on different subjects, meaning that 
a subject's data will appear exclusively in either the training or the test set. 
This task evaluates the model's adaptability to unseen individuals, which is crucial for its 
viability in large-scale deployment scenarios.

For performance evaluation, we select the \textbf{F1-score}  as our primary metric. 
Given that the F1-score balances precision and recall, 
it provides a robust assessment even in the presence of data imbalance. Following our validation protocol, 
the model checkpoint achieving the highest F1-score on the validation set is chosen 
for final evaluation. We then report its F1-score on the test set. Furthermore, we also 
report \textbf{accuracy}, as it is a key metric for classification tasks.

\section{Experiments}
\label{sec:experiment}
We first reproduce the model by strictly following the settings from the source 
paper. We consider the reproduction successful if its performance is within 10 \%
of the originally reported results and incorporate it into our algorithm library. 
These validated models are then used for further experiments on our defined 
benchmark, followed by a complete evaluation and analysis of the results.
\subsection{Model Reproduction}
\label{ssec:reproduction}

\begin{table}[t!]
  \centering
  \caption{Results of reproduced models in LibEMER}
  \label{tab:reproduction}
  \footnotesize
  \setlength\tabcolsep{2.5pt}

  \begin{tblr}{
    width = \columnwidth,
    colspec = {X[1.6, l, m] X[1.1, l, m] X[0.5, c, m] X[0.9, c, m] X[0.9, c, m] X[0.7, c, m]},
    rowsep = 0.5pt
  }
  \hline
  \SetCell[r=2]{m} Method & \SetCell[c=2]{c} \makecell{Experimental\\Setup} & & \SetCell[c=3]{c} \makecell{Results\\(Accuracy)} & & \\
  \cline{2-3} \cline{4-6}
  & Dataset & Task & Reported & Ours & Gap \\
  \hline
  \SetCell[r=3]{m} DCCA & SEED & SD & 94.58 & 86.50 & \showgap{94.58}{86.50}\\
  \cline{2-6}
   & DEAP-V & SD & 85.62 & 87.49 & \showgap{85.62}{87.49} \\
  \cline{2-6}
   & DEAP-A & SD & 84.33 & 89.09 & \showgap{84.33}{89.09} \\
  \hline
  DCCA\_AM & SEEDV & SD & 85.30 & 76.84 & \showgap{85.30}{76.84} \\
  \hline
  \makecell[l]{Bimodal\\LSTM} & SEED & SD & 93.97 & 84.21 & \showgap{93.97}{84.21} \\
  \hline
  BDDAE & SEED & SD & 88.98 & 80.37 & \showgap{88.98}{80.37} \\
  \hline
  MCAF & SEED & SD & 96.26 & 88.32 & \showgap{96.26}{88.32} \\
  \hline
  G2G & SEEDV & SD & 83.87 & 82.21 & \showgap{83.87}{82.21} \\
  \hline
  \makecell[l]{CFDA-\\CSF} & SEEDV & SI & 80.01 & 74.74 & \showgap{80.01}{74.74} \\
  \hline
  \SetCell[r=2]{m} CMCM & DEAP-V & SD & 79.62 & 81.51 & \showgap{79.62}{81.51} \\
  \cline{2-6}
   & DEAP-A & SD & 81.69 & 82.10 & \showgap{81.69}{82.10} \\
  \hline
  \makecell[l]{HetEmotion\\Net} & DEAP-V & SD & 97.66 & 88.17 & \showgap{97.66}{88.17} \\
  \hline
  CRNN & DEAP-V & SD & 91.05 & 87.05 & \showgap{91.05}{87.05} \\
  \hline
   \end{tblr}
\end{table}
As Table\ref{tab:reproduction} shows, the performance of reproduced methods, when compared to their originally reported values, 
exhibited a decline across most tasks. However, these performance discrepancies were consistently 
within a 10\% margin. Conversely, a minority of algorithms showed marginal performance gains on specific tasks.
These variations can be primarily attributed to several pervasive challenges. A principal 
factor is the insufficiently detailed description of model architectures\cite{tang2017multimodal,yin2024research}, data preprocessing pipelines, 
and experimental settings\cite{liao2020multimodal} within the source publications. Even in cases where code was provided\cite{jia2021hetemotionnet,jin2023graph,liu2021comparing}, 
significant performance deviations arose from unavoidable disparities in hardware environments, 
software platforms, and library versions.
\subsection{Result and Analysis}
\label{ssec:result}

\begin{table*}[t] 
  \centering
  \caption{The mean and standard deviations of accuracy (\%) and F1-score (\%) for chosen methods following the proposed benchmark on subject-dependent and subject-independent tasks.
  The best model is highlighted in bold. The second best is underlined.}
  \label{tab:result}
\scriptsize

\setlength\tabcolsep{2.5pt} 

\begin{tblr}{
  width = \linewidth, 
  colspec = {Q[l, m] *{16}{X[c, m]}}, 
  stretch = 0,
  cell{1}{1} = {r=3}{},
  cell{1}{2} = {c=8}{},
  cell{1}{10} = {c=8}{},
  cell{2}{2} = {c=2}{},
  cell{2}{4} = {c=2}{},
  cell{2}{6} = {c=2}{},
  cell{2}{8} = {c=2}{},
  cell{2}{10} = {c=2}{},
  cell{2}{12} = {c=2}{},
  cell{2}{14} = {c=2}{},
  cell{2}{16} = {c=2}{},
  hline{1,4,14} = {-}{},
  hline{3} = {2-17}{lr},  
  rowsep = 1.5pt
}
Method        & Subject-dependent &       &       &       &        &       &        &       & Subject-independent &       &       &       &        &       &        &       \\
\cline{2-9} \cline{10-17}
              & SEED         &       & SEEDV &       & DEAP-V &       & DEAP-A &       & SEED           &       & SEEDV &       & DEAP-V &       & DEAP-A &       \\
              & ACC          & F1    & ACC   & F1    & ACC    & F1    & ACC    & F1    & ACC            & F1    & ACC   & F1    & ACC    & F1    & ACC    & F1    \\
CFDA-CSF      & {\underline{81.03}\\\underline{(20.24)}} & {\underline{75.85}\\\underline{(25.41)}} & {22.85\\(23.09)} & {\underline{15.77}\\\underline{(15.80)}} & {55.57\\(14.79)}  & {49.35\\(14.82)} & {55.13\\(19.49)}  & {45.65\\(17.88)} & {38.06\\(0.20)}         & {32.72\\(4.80)} & {\textbf{45.57}\\\textbf{(23.64)}} & {\textbf{46.44}\\\textbf{(19.12)}} & {52.41\\(9.41)}  & {\underline{45.84}\\\underline{(12.64)}} & {50.83\\(11.91)}  & {\textbf{50.06}\\\textbf{(4.96)}} \\
\hline
BDDAE         & {79.23\\(19.46)}        & {74.07\\(24.41)} & {16.94\\(21.99)} & {11.30\\(14.79)} & {55.60\\(16.88)}  & {46.84\\(17.26)} & {58.90\\(14.93)}  & {45.91\\(12.86)} & {\textbf{56.89}\\\textbf{(6.89)}}         & {\textbf{48.92}\\\textbf{(0.26)}} & {25.11\\(8.52)} & {22.86\\(11.93)} & {\textbf{55.83}\\\textbf{(5.16)}}  &{35.77\\(1.95)} & {\textbf{54.58}\\\textbf{(9.14)}}  & {35.12\\(3.46)} \\
\hline
CMCM          & {78.30\\(20.92)}        & {72.08\\(20.26)} & {13.97\\(24.10)} & {8.77\\(15.94)}  & {55.90\\(12.98)}  & {\textbf{55.75}\\\textbf{(5.11)}} & {59.14\\(16.41)}  & {\textbf{52.19}\\\textbf{(14.10)}} & {26.31\\(8.60)}          & {15.58\\(1.94)} & {\underline{42.92}\\\underline{(12.47)}} & {\underline{43.15}\\\underline{(13.05)}} & {\underline{55.75}\\\underline{(5.11)}}  & {\textbf{49.84}\\\textbf{(5.25)}} & {47.61\\(5.47)}  & {\underline{46.61}\\\underline{(6.29)}} \\
\hline
DCCA          & {71.88\\(25.11)}        & {66.10\\(29.38)} & {19.24\\(24.14)} & {14.15\\(19.40)} & {\underline{59.12}\\\underline{(12.46)}}  & {51.57\\(14.29)} & {57.23\\(16.45)}  & {46.78\\(13.70)} & {33.91\\(22.42)}        & {27.19\\(12.82)} & {23.40\\(8.24)} & {13.68\\(6.98)} & {52.61\\(8.50)}  & {36.80\\(1.91)} & {49.58\\(10.66)}  & {35.85\\(6.90)} \\
\hline
DCCA\_AM      & {\textbf{82.01}\\\textbf{(19.72)}}        & {\textbf{77.64}\\\textbf{(23.87)}} & {\textbf{35.38}\\\textbf{(27.30)}} & {\textbf{23.84}\\\textbf{(21.10)}} & {52.41\\(15.71)}  & {48.70\\(16.74)} & {58.18\\(14.70)}   & {\underline{51.04}\\\underline{(13.96)}} & {\underline{44.00}\\\underline{(22.59)}}      & {\underline{37.17}\\\underline{(20.88)}} & {32.65\\(17.85)} & {21.50\\(18.92)} & {55.24\\(8.67)}  & {45.78\\(11.02)} & {46.99\\(7.67)}  & {41.77\\(6.73)} \\
\hline
CRNN          & {59.56\\(18.77)}        & {49.05\\(22.34)}   & {14.00\\(19.57)} & {10.02\\(14.13)} & {\textbf{59.36}\\\textbf{(9.84)}}  & {\underline{53.12}\\\underline{(11.49)}} &{58.43\\(18.00)}  & {49.80\\(16.18)} & {38.84\\(8.90)}          & {27.69\\(11.32)} & {39.40\\(9.69)} & {38.92\\(6.68)} & {45.28\\(4.79)}  & {32.81\\(4.64)} & {\underline{53.39}\\\underline{(9.90)}}  & {36.91\\(5.14)} \\
\hline
BimodalLSTM   & {66.99\\(31.98)}        & {61.83\\(33.50)} & {16.12\\(25.31)} & {9.68\\(16.08)}  & {53.67\\(15.58)}  & {47.07\\(16.09)} & {\underline{59.19}\\\underline{(18.92)}} & {48.85\\(19.67)} & {43.48\\(15.58)}          & {34.62\\(18.28)} & {38.45\\(8.33)} & {31.81\\(6.53)}  & {53.56\\(7.51)}  & {36.54\\(4.05)} & {52.22\\(8.74)}  & {42.61\\(7.48)} \\
\hline
MCAF          & {52.11\\(24.77)}        & {42.75\\(27.71)} & {7.64\\(16.39)}  & {4.80\\(9.83)}  & {53.59\\(16.34)}  & {46.42\\(16.38)} & {\textbf{62.29}\\\textbf{(15.21)}}  & {48.63\\(14.09)} & {29.45\\(5.37)}          & {22.72\\(9.11)} & {22.71\\(2.57)} & {11.99\\(3.26)} & {52.08\\(7.49)}  & {36.34\\(5.44)} & {48.76\\(8.37)}  & {37.16\\(8.27)} \\
\hline
G2G           & {44.61\\(17.19)}        & {37.12\\(17.70)} & {\underline{26.45}\\\underline{(23.35)}} & {15.33\\(15.72)}     & {54.69\\(8.17)}      & {50.15\\(9.67)}     & {56.60\\(10.94)}      & {50.60\\(8.52)}     & {32.90\\(0.22)}    & {21.25\\(4.34)}     & {39.42\\(3.24)}     & {34.75\\(1.71)}     & {51.80\\(6.99)}      & {40.84\\(5.71)}     & {42.92\\(6.32)}      & {33.41\\(3.36)}     \\
\hline
HetEmotionNet & -            & -     & -     & -     & {56.22\\(6.87)}      & {50.52\\(8.53)}     & {53.66\\(11.45)}      & {46.59\\(8.67)}     & -              & -     & -     & -     & {53.2\\(4.38)}      & {51.27\\(6.22)}     & {48.84\\(7.49)}      & {41.21\\(8.86)}     
\end{tblr}
\end{table*}

Table\ref{tab:result} shows the results from evaluating the models under a unified framework with standardized preprocessing and a three-way split. 
Notably, HetEmotionNet is tested only on DEAP due to its specific architecture. We derive the following key analyses.
\vspace{0.5em}

\noindent\textbf{Observation 1: Physiological signals provide an objective reflection of responses to external stimuli,aligning more closely with objective, discrete labels than with subjective ratings.
Objective data with clear boundaries serve as a higher-quality benchmark for robust evaluation.}

Models generally show poor performance on the binary classification tasks of the DEAP dataset, whereas they perform better overall on the 
three-class classification tasks of the SEED dataset. We attribute this to label evaluation and modality characteristics. 
SEED pre-screens the target emotions for its stimuli to obtain discrete and objective labels. In contrast, the DEAP 
dataset uses participants' subjective ratings on the two continuous dimensions of arousal and valence as labels. Although peripheral 
physiological signals can objectively reflect physiological states under different stimuli, they 
are not controlled by the central nervous system and are less effective for recognizing subjective cognition. 
Simultaneously, learning complex thought patterns unrelated to emotional stimuli from EEG data is also more challenging. 
Therefore, objective and discrete labels are clean and suitable for classification tasks.

\begin{figure}[htb]
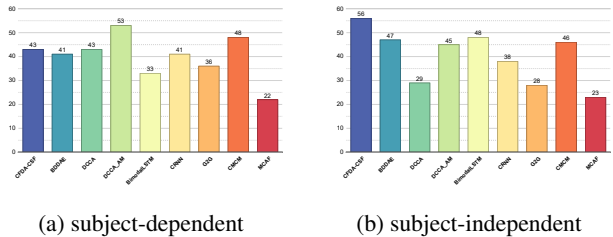

  \begin{minipage}[b]{0.5\linewidth}
  \centering
 \centerline{\includegraphics[width=4.0cm]{figures/dependent\_score.pdf}}
  \centerline{(a) subject-dependent}
  \end{minipage}
  \hfill
  \begin{minipage}[b]{0.5\linewidth}
  \centering
  \centerline{\includegraphics[width=4.0cm]{figures/independent\_score.pdf}}
  \centerline{(b) subject-independent}
  \end{minipage}
  \caption{Scores of model performance on two tasks based on Accuracy and F1-score}
  \label{fig:score}
  \vspace{-3mm}
  \end{figure}
\vspace{0.5em} 
\noindent\textbf{Observation 2: The strategic design of learning tasks and fusion strategies tailored to modal heterogeneity is more critical for performance enhancement than the choice of model architecture alone.}

To evaluate this, we ranked the models that completed all experiments on a scale from n to 1 based on two key metrics for each dataset. The final score for each model 
is the aggregate of its rankings across all metrics. Figure\ref{fig:score} shows scores on two tasks. In subject-dependent tasks, DCCA\_AM demonstrated the best overall performance, 
whereas CFDA-CSF achieved the highest score in subject-independent tasks. Concurrently, DNN-based methods presented more robust performance 
across both task categories, with three such models securing the top rank in 11 out of 16 metrics. The success is mainly attributed to
their meticulously designed learning tasks and strategies for cross-modal representation, which facilitate the alignment of modal distributions
to improve cross-modal correlation and fusion reliability. Within the Transformer-based models, CMCM significantly outperformed MCAF. 
This disparity arises because CMCM addresses modal heterogeneity through a credibility-fusion strategy, exploiting a cross-attention mechanism to 
model sequence consistency. In contrast, MCAF primarily focuses on channel selection and enhancing representation. 
CNN-based methods showed relatively weaker performance. However, models incorporating RNN architectures performed well on certain tasks, 
suggesting that their inherent sequence learning capabilities hold considerable potential. Based on sophisticated cross-modal strategy , architectures capable of modeling temporal or 
spatial properties may further achieve more effective representation learning.

\vspace{0.5em}
\noindent\textbf{Observation 3: The sparsity constrains the representational capacity of deep learning models, 
hindering their ability to learn complex cross-modal emotion representations.}

A significant performance discrepancy was observed under the unified benchmark, with all models 
performing substantially below their originally reported results. This is further underscored by the large standard deviations across all tasks, 
which in some cases exceeded the mean, indicating high performance instability. These findings highlight that class and inter-subject variability 
remain formidable challenges. Deep learning methods rely on large-scale, high-quality data to adequately learn complex patterns. However, 
the limited size of mainstream dataset with the absence of a validation set in many experimental settings, intensifies the risk of 
overfitting and leads to inflated performance claims. Our rigorous setup reveals that existing models hard to effectively 
learn complex cross-modal emotion representations from such sparse data. Therefore, the establishment of large-scale, 
high-quality datasets with a greater diversity of stimuli and subjects is imperative for advancing the development of EMER.

\section{CONCLUSION}
\label{sec:conclusion}
In this paper, we proposed a unified benchmark and algorithm library for EEG-based multimodal emotion recognition named LibEMER.
LibEMER includes 10 models and standardize end-to-end workflow to support unbiased evaluation across three most popular public datasets
on two primary tasks. We put forward some key findings and analyses and identify current research challenges and directions through fair experiments to inspire research in related fields.
We believe that LibEMER will accelerate the standardization and transparency process and facilitate the development of new algorithms.
In the future, we plan to continue to improve the library and expand the scope of the benchmark.

\vfill\pagebreak

\section{ACKNOWLEDGEMENT}
This work was supported by the National Natural Science
Foundation of China (62202367), Project of China
Knowledge Centre for Engineering Science and Technology.


\bibliographystyle{IEEEbib}
\bibliography{reference}

\begin{thebibliography}{10}

\bibitem{zhang2024torcheegemo}
Zhi Zhang, Sheng-hua Zhong, and Yan Liu,
\newblock ``Torcheegemo: A deep learning toolbox towards eeg-based emotion recognition,''
\newblock {\em Expert Systems with Applications}, vol. 249, pp. 123550, 2024.

\bibitem{zhang2024gnn4eeg}
Kaiyuan Zhang, Ziyi Ye, Qingyao Ai, Xiaohui Xie, and Yiqun Liu,
\newblock ``Gnn4eeg: A benchmark and toolkit for electroencephalography classification with graph neural network,''
\newblock in {\em Companion of the 2024 on ACM International Joint Conference on Pervasive and Ubiquitous Computing}, 2024, pp. 612--617.

\bibitem{liu2025libeer}
Huan Liu, Shusen Yang, Yuzhe Zhang, Mengze Wang, Fanyu Gong, Chengxi Xie, Guanjian Liu, Zejun Liu, Yong-Jin Liu, Bao-Liang Lu, et~al.,
\newblock ``Libeer: A comprehensive benchmark and algorithm library for eeg-based emotion recognition,''
\newblock {\em IEEE Transactions on Affective Computing}, 2025.

\bibitem{paszke2019pytorch}
Adam Paszke, Sam Gross, Francisco Massa, Adam Lerer, James Bradbury, Gregory Chanan, Trevor Killeen, Zeming Lin, Natalia Gimelshein, Luca Antiga, et~al.,
\newblock ``Pytorch: An imperative style, high-performance deep learning library,''
\newblock {\em Advances in neural information processing systems}, vol. 32, 2019.

\bibitem{abadi2016tensorflow}
Mart{\'\i}n Abadi, Paul Barham, Jianmin Chen, Zhifeng Chen, Andy Davis, Jeffrey Dean, Matthieu Devin, Sanjay Ghemawat, Geoffrey Irving, Michael Isard, et~al.,
\newblock ``$\{$TensorFlow$\}$: a system for $\{$Large-Scale$\}$ machine learning,''
\newblock in {\em 12th USENIX symposium on operating systems design and implementation (OSDI 16)}, 2016, pp. 265--283.

\bibitem{liu2024eeg}
Huan Liu, Tianyu Lou, Yuzhe Zhang, Yixiao Wu, Yang Xiao, Christian~S Jensen, and Dalin Zhang,
\newblock ``Eeg-based multimodal emotion recognition: A machine learning perspective,''
\newblock {\em IEEE Transactions on Instrumentation and Measurement}, vol. 73, pp. 1--29, 2024.

\bibitem{pillalamarri2025review}
Rajasekhar Pillalamarri and Udhayakumar Shanmugam,
\newblock ``A review on eeg-based multimodal learning for emotion recognition,''
\newblock {\em Artificial Intelligence Review}, vol. 58, no. 5, pp. 131, 2025.

\bibitem{zheng2015investigating}
Wei-Long Zheng and Bao-Liang Lu,
\newblock ``Investigating critical frequency bands and channels for eeg-based emotion recognition with deep neural networks,''
\newblock {\em IEEE Transactions on autonomous mental development}, vol. 7, no. 3, pp. 162--175, 2015.

\bibitem{li2019classification}
Tian-Hao Li, Wei Liu, Wei-Long Zheng, and Bao-Liang Lu,
\newblock ``Classification of five emotions from eeg and eye movement signals: Discrimination ability and stability over time,''
\newblock in {\em 2019 9th International IEEE/EMBS Conference on Neural Engineering (NER)}. IEEE, 2019, pp. 607--610.

\bibitem{koelstra2011deap}
Sander Koelstra, Christian Muhl, Mohammad Soleymani, Jong-Seok Lee, Ashkan Yazdani, Touradj Ebrahimi, Thierry Pun, Anton Nijholt, and Ioannis Patras,
\newblock ``Deap: A database for emotion analysis; using physiological signals,''
\newblock {\em IEEE transactions on affective computing}, vol. 3, no. 1, pp. 18--31, 2011.

\bibitem{qiu2018multi}
Jie-Lin Qiu, Wei Liu, and Bao-Liang Lu,
\newblock ``Multi-view emotion recognition using deep canonical correlation analysis,''
\newblock in {\em International conference on neural information processing}. Springer, 2018, pp. 221--231.

\bibitem{liu2021comparing}
Wei Liu, Jie-Lin Qiu, Wei-Long Zheng, and Bao-Liang Lu,
\newblock ``Comparing recognition performance and robustness of multimodal deep learning models for multimodal emotion recognition,''
\newblock {\em IEEE Transactions on Cognitive and Developmental Systems}, vol. 14, no. 2, pp. 715--729, 2021.

\bibitem{tang2017multimodal}
Hao Tang, Wei Liu, Wei-Long Zheng, and Bao-Liang Lu,
\newblock ``Multimodal emotion recognition using deep neural networks,''
\newblock in {\em International Conference on Neural Information Processing}. Springer, 2017, pp. 811--819.

\bibitem{jimenez2024cfda}
Magdiel Jim{\'e}nez-Guarneros and Gibran Fuentes-Pineda,
\newblock ``Cfda-csf: A multi-modal domain adaptation method for cross-subject emotion recognition,''
\newblock {\em IEEE Transactions on Affective Computing}, vol. 15, no. 3, pp. 1502--1513, 2024.

\bibitem{jin2023graph}
Ming Jin and Jinpeng Li,
\newblock ``Graph to grid: Learning deep representations for multimodal emotion recognition,''
\newblock in {\em Proceedings of the 31st ACM International Conference on Multimedia}, 2023, pp. 5985--5993.

\bibitem{liao2020multimodal}
Jinxiang Liao, Qinghua Zhong, Yongsheng Zhu, and Dongli Cai,
\newblock ``Multimodal physiological signal emotion recognition based on convolutional recurrent neural network,''
\newblock in {\em IOP conference series: materials science and engineering}. IOP Publishing, 2020, vol. 782, p. 032005.

\bibitem{zhang2024cross}
Yuzhe Zhang, Huan Liu, Di~Wang, Dalin Zhang, Tianyu Lou, Qinghua Zheng, and Chai Quek,
\newblock ``Cross-modal credibility modelling for eeg-based multimodal emotion recognition,''
\newblock {\em Journal of Neural Engineering}, vol. 21, no. 2, pp. 026040, 2024.

\bibitem{yin2024research}
Jialai Yin, Minchao Wu, Yan Yang, Ping Li, Fan Li, Wen Liang, and Zhao Lv,
\newblock ``Research on multimodal emotion recognition based on fusion of electroencephalogram and electrooculography,''
\newblock {\em IEEE Transactions on Instrumentation and Measurement}, vol. 73, pp. 1--12, 2024.

\bibitem{jia2021hetemotionnet}
Ziyu Jia, Youfang Lin, Jing Wang, Zhiyang Feng, Xiangheng Xie, and Caijie Chen,
\newblock ``Hetemotionnet: two-stream heterogeneous graph recurrent neural network for multi-modal emotion recognition,''
\newblock in {\em Proceedings of the 29th ACM international conference on multimedia}, 2021, pp. 1047--1056.

\bibitem{duan2013differential}
Ruo-Nan Duan, Jia-Yi Zhu, and Bao-Liang Lu,
\newblock ``Differential entropy feature for eeg-based emotion classification,''
\newblock in {\em 2013 6th international IEEE/EMBS conference on neural engineering (NER)}. IEEE, 2013, pp. 81--84.

\bibitem{sanei2013eeg}
Saeid Sanei and Jonathon~A Chambers,
\newblock {\em EEG signal processing},
\newblock John Wiley \& Sons, 2013.

\bibitem{duan2012eeg}
Ruo-Nan Duan, Xiao-Wei Wang, and Bao-Liang Lu,
\newblock ``Eeg-based emotion recognition in listening music by using support vector machine and linear dynamic system,''
\newblock in {\em International conference on neural information processing}. Springer, 2012, pp. 468--475.

\bibitem{lu2015combining}
Yifei Lu, Wei-Long Zheng, Binbin Li, and Bao-Liang Lu,
\newblock ``Combining eye movements and eeg to enhance emotion recognition.,''
\newblock in {\em IJCAI}. Buenos Aires, 2015, vol.~15, pp. 1170--1176.

\end{thebibliography}

\end{document}